\documentclass[10pt]{article}
\usepackage{amsmath,amsfonts}
\usepackage{multirow}
\usepackage{amssymb}
\usepackage[colorlinks=true,urlcolor=blue]{hyperref}
\usepackage{graphicx}
\usepackage{nicefrac}
\usepackage{ulem}
\usepackage{cancel}
\hypersetup{                
backref = true,             
pagebackref = true,                 
hyperindex = true,          
colorlinks = true,          
breaklinks = true,          
urlcolor = blue,            
linkcolor = blue,           
bookmarks = true,           
bookmarksopen = true,       
citecolor=red,
}

\newcommand{\bpm}{\begin{pmatrix}}
\newcommand{\epm}{\end{pmatrix}}
\newcommand{\qeed}{\hfill\textrm{QED}\break\null}

\advance \textheight by \topskip
\makeatletter

\def\cc#1{\kern .7em\hfill #1 \hfill\kern .7em}

\newcommand{\bb}{\color{black}}

\newcommand{\beqa}{\begin{eqnarray}}
\newcommand{\eeqa}{\end{eqnarray}}

\newcommand{\nn}{\nonumber}

\newcommand{\ii}{{\mathrm i}}
\newcommand{\g}{{\mathfrak{g}}}
\newcommand{\no}{{}_\circ \hskip -.17truecm {}^\circ \hskip .1 truecm}
\newcommand{\noo}{{}_{\times} \hskip -.235 truecm  {}^{\times}}

\title{Fermion realisations of generalised Kac--Moody   and Virasoro algebras  associated to the two-sphere and the two-torus}
\begin{document} 
\maketitle
\begin{center}
Rutwig Campoamor-Stursberg$^{1\ast}$, Michel Rausch de Traubenberg$^{2\dagger}$
\medskip

$^1$  Instituto de Matem\'atica Interdisciplinar and Dpto. Geometr\'\i a y Topolog\'\i a, UCM, E-28040 Madrid, Spain\\

$^3$ Universit\'e de Strasbourg, CNRS, IPHC UMR7178, F-67037 Strasbourg Cedex, France\\  

\end{center}
  \noindent $^\ast$ Email: rutwig@ucm.es

\noindent $^\dagger$ Email:  Michel.Rausch@iphc.cnrs.fr 

\medskip

\begin{abstract}
 Using the notion of extension of Kac-Moody algebras for higher dimensional compact manifolds  recently introduced, we show that for the two-torus $\mathbb S^1 \times \mathbb S^1$ and the two-sphere $\mathbb S^2$,  these extensions, as well as extensions of the Virasoro algebra can be  obtained naturally from the usual Kac-Moody and Virasoro algebras. Explicit fermionic realisations  are proposed.  In order to have well defined generators, beyond the usual normal ordering prescription, we introduce a regulator and regularise infinite sums by means of Riemann $\zeta-$function.

{\bf keywords}:  Kac-Moody; Virasoro algebras; fermion realization; regularisation
\end{abstract}



 In the context of generalized Kac--Moody algebras, a new type of infinite dimensional Lie algebras associated to compact real manifolds has recently been introduced and studied in \cite{rmm}. As Kac-Moody and Virasoro algebras are related to the one-dimensional compact manifold $\mathbb S^1$, it can naturally be inferred that these generalized structures can be related with higher-dimensional real compact manifolds ${\cal M}$,  where ${\cal M}$ is either a compact Lie group $G_c$ or  a compact homogeneous space  $ G_c/H$, with $H$ being a closed (not necessarily maximal) subgroup of $G_c$.
Central extensions and  the representation theory of such algebras were analyzed,
associating them to the properties
and representations of $G_c$ or ${\cal M}$.   From a slightly different perspective, such extensions of Kac-Moody algebras have also  been considered in \cite{bars, KT,Frap} for ${\cal M} = \mathbb S^2$ and ${\cal M}= \mathbb S^1\times \cdots \times \mathbb S^1$, and more recently in \cite{multil, jap} (where the algebra related to $\mathbb S^1\times \cdots \times \mathbb S^1$ is referred to as a multi-loop algebra).

In this note we construct an explicit fermionic realisation for extensions of the Kac-Moody and the Virasoro algebras on the two-torus $\mathbb S^1 \times \mathbb S^1$ and the two-sphere $\mathbb S^2$, and show that these extensions can be obtained in a natural manner from the usual Kac-Moody and Virasoro algebras.

 Beginning with the two-sphere, we observe that with $u=\cos \theta$, the usual 
spherical harmonics can be written as
\beqa
\label{eq:Y}
Y_{\ell m}(u,\varphi)= \sqrt{\frac{(\ell -m)!}{(\ell+m)!}}P_{\ell m}(u)e^{\ii m \varphi} = Q_{\ell m}(u) e^{\ii m \varphi}\nn
\eeqa
where $P_{\ell m}$ are the associated Legendre functions \cite{be}. Hence, for elements $\Phi$ in the Hilbert space $L^2(\mathbb S^2)$ we can write
\beqa
\Phi(u,\varphi) = \sum\limits_{m=-\infty}^{+ \infty} \Bigg( \sum \limits_{\ell = |m|}^{+ \infty} \Phi^{\ell,m}  Q_{\ell m}(u)\Bigg)
                         e^{\ii m\varphi}
                         = \sum\limits_{m=-\infty}^{+ \infty} \Phi_m(u) e^{\ii m \varphi} \ , \nn
\eeqa
where
\beqa
\label{eq:expP}
\Phi^{\ell,m}&=&(Q_{\ell m}, \Phi_m)=\frac 12 \int \limits_{-1}^1  \text{d} u \;Q_{\ell m}(u) \Phi_m(u) \ ,\nn\\
            &=&(Y_{\ell m}, \Phi) = \frac 1 {4 \pi}  \int \limits_{0}^{2\pi}  \text{d} \varphi \int \limits_{-1}^1  \text{d} u \;\bar Y^{\ell m}(u,\varphi) \Phi(u, \varphi) \ .\nn
\eeqa
 It follows in particular that
\beqa
\label{eq:Bn}
{\cal B}_m =\big\{Q_{\ell m}, \ell\ge |m|\big\} \ , \  \ \forall m \in \mathbb Z \ 
\eeqa
forms an orthonormal  Hilbert basis and we have
\beqa
\label{eq:delS2R}
\delta(u-v) = \sum \limits_{\ell \ge |m|} Q_{\ell m}(u) Q_{\ell m}(v) \ . 
\eeqa
Consider now a compact Lie algebra $\g$ with structure constants $f^{ab}{}_c$ and introduce the usual Kac-Moody and Virasoro algebras
generated by $T^a_m, L_m$. Suppose further that the generators of the Virasoro and Kac-Moody algebras depend on the  variable $u$.  Using the basis \eqref{eq:Bn}, we can decompose these generators as 
\beqa
T_m^a(u) = \sum \limits_{\ell =|m|}^{+ \infty} T_{\ell,m}^a Q_{\ell m}(u) \ , \ \ 
L_m(u)=&\sum \limits_{\ell =|m|}^{+ \infty} L_{\ell,m} P_{\ell m}(u) \ . \nn\
\eeqa
We assume further the following Lie brackets  (note that Jacobi identities are automatically satisfied)
\beqa
\big[L_m(u),L_n(v)\big]&=&(m-n) L_{m+n} (v)\delta(u-v)+ \frac c{12}m(m^2-1) \delta_{m+n}\delta(u-v) \nn\\
\big[T_{m_1}^{a_1}(u), T_{m_2}^{a_2}(v) \big] &=& \ii f^{a_1 a_2}{}_{a_3} T^{a_3}_{m_1+m_2}(v) \delta(u-v)+ k m_1 \delta^{a_1 a_2} \delta_{m_1+m_2}\delta(u-v) \  \nn\\
\big[L_m(u),T_{n}^{a}(v)\big] &=&-n  T^a_{m+n} (v)\delta(u-v)\ , \nn 
\eeqa
thus extracting the modes on both sides of the equations above we obtain
\beqa
\label{eq:mod}
\big[L_{\ell_1,m_1},L_{\ell_2,m_2}\big]&=&(m_1-m_2)\; c_{\ell_1,m_1,\ell_2,m_2}{}^{\ell_3,m_1+m_2} \; L_{\ell_3,m_1+m_2} \nn\\
&&+(-1)^{m_1} \frac c{12}m_1(m_1^2-1)
\delta_{m_1+m_2}\delta_{\ell_1\ell_2} \nn\\
\big[T_{\ell_1,m_1}^{a_1}, T_{\ell_2,m_2}^{a_2}\big]&=&\ii f^{a_1a_2}{}_{a_3} \; c_{\ell_1,m_1,\ell_2,m_2}{}^{\ell_3,m_1+m_2} \;T^{a_3}_{\ell_3,m_1+m_2}\nn\\
&&+(-1)^{m_1} k m_1 \delta_{m_1+m_2}\delta_{\ell_1\ell_2} \\
\big[L_{\ell_1,m_1},T_{\ell_2,m_2}^{a_2}\big]&=&-m_2 \;c_{\ell_1,m_1,\ell_2,m_2}{}^{\ell_3,m_1+m_2} T^{a_2}_{\ell_3,m_1+m_2}\ , \nn
\eeqa
where the coefficients $c_{\ell_1,m_1,\ell_2,m_2}{}^{\ell_3,m_1+m_2}$ are related to the Clebsch-Gordan coefficients  (for explicit expressions see  \cite{rmm}). 
This construction enables us to obtain in a natural way the Kac-Moody and Virasoro algebras associated to the sphere $\mathbb S^2$
(defined in \cite{rmm} for the Kac-Moody algebra). 
 Note that the derivative operator $d$ introduced in \cite{rmm}  coincides with $-L_{0,0}$.
The last bracket in (\ref{eq:mod}) explicitly shows the semi-direct structure of both algebras.   It is worthy to be observed that
$\{L_{\ell,m}, m\in \mathbb Z, \ell\ge |m|\}$ generates a subalgebra of vector fields on $\mathbb S^2$. This algebra, as seen in the first equation in \eqref{eq:mod}, clearly admits a central
extension.
 It is well known that to centrally extend a Lie algebra various difficulties have to be surmounted, as the Jacobi identities must be satisfied and  the central term (charge) has to be non-trivial, {\it i.e.}, should not vanish by a change of basis. In the case of the semi-direct structure above, the situation is even more difficult as the centrally extended subalgebra of vector fields on
$\mathbb S^2$ has to be compatible with the Kac-Moody algebra associated to $\mathbb S^2$.   Centrally extended Lie algebras of vector fields on $\mathbb S^2$ have been studied in \cite{FI}, while the compatibility  condition of centrally extended algebras of vector field of the two-sphere and the corresponding Kac-Moody algebra has been studied in \cite{Frap,RS,RS2}.

\medskip
The case of the torus $\mathbb S^1 \times \mathbb S^1$ is structurally simpler.
Let $(\varphi_1,\varphi_2)$ be the two angles of the two-torus. Assume that we have the usual Kac-Moody and Virasoro algebra associated to one direction of the torus, say to $\varphi_1$. Denote
$L_m, T^a_m$ the corresponding generators. Then, as in the case of the two-sphere, we assume that these generators depend on the variable $\varphi_2$ 
\beqa
\label{eq:TKM}
T_m^a(\varphi_2) = \sum \limits_{n=-\infty}^{+ \infty} T_{m,n}^a e^{-\ii n \varphi_2} \ , \ \ 
L_m(\varphi_2) = \sum \limits_{n=-\infty}^{+ \infty} L_{m,n}  e^{-\ii n \varphi_2}.  \
\eeqa
By analogous considerations as those made for the two-sphere, we can derive the Lie brackets 
\beqa
\big[L_{m_1,n_1},L_{m_2,n_2}\big] &=& (m_1-m_2) L_{m_1+m_2,n_1+n_2} + \frac {c}{12}m_1(m_1^2-1)\delta_{m_1+m_2}\delta_{n_1+n_2}\nn\\
\big[T^{a_1}_{m_1,n_1},T^{a_2}_{m_2,n_2}\big] &=& \ii f^{a_1 a_2}{}_{a_3} T^{a_2}_{m_1+m_2,n_1+n_2} + k \delta^{ab} m_1 \delta_{m_1+m_2}\delta_{n_1+n_2}\nn\\
\big[L_{m_1,n_1},T^{a}_{m_2,n_2}\big]&=&-m_2 T^a_{m_1+m_2,n_1+n_2} \ .\nn
\eeqa
As for the two-sphere, the derivative operator $d$ associated to the Kac-Moody algebra introduced in \cite{rmm} coincides with $-L_{0,0}$.
  We further observe that the subalgebra $\{L_{m,0},c, m\in \mathbb N\}\ltimes
\{T_{m,0}^a,k, a=1,\cdots,\dim \g, m \in \mathbb Z\}$ is isomorphic to the semi-direct product of the Virasoro algebra with the usual affine Lie algebra. Note also the dissymmetry between the two variables $(\varphi_1,\varphi_2)$, which is a consequence of our construction procedure
 of the algebra associated to $\mathbb S^1\times \mathbb S^1$ and, in particular, of our ordering relation (see below \eqref{eq:ord}). In principle, it is conceivable to construct a more symmetric version of the algebra associated to the two-torus (in the Kac-Moody case it corresponds to an extension with two central extensions as shown in
\cite{rmm}). These symmetric extensions will  however not be considered in this short note.\\

Now we turn to the fermion realisation of these algebras.
Consider first  the case of the two-torus parameterised by the variables $(\varphi_1,\varphi_2)\in [0,2 \pi[ \times [0,2 \pi[ $ and introduce $d$ real fermions $H^i$ on $\mathbb S^1 \times \mathbb S^1$.
With the usual conformal algebra techniques in mind, we fix one direction of the
two-torus and write (this in fact corresponds to the  Wick rotation  $\exp(\ii \varphi_1)\to z \in \mathbb C$, the other direction
being noted $\varphi$)
\beqa
H^i(z,\varphi)= \sum \limits_{m\in \mathbb Z (+\frac12)} b^i_m(\varphi) z^{-m}= \sum  \limits_{m\in \mathbb Z (+\frac12)}
\sum  \limits_{p\in \mathbb Z (+\frac12)} b_{mp}^i z^{-m}  e^{- \ii p\varphi}\ ,
  \ \ i=1,\cdots, d \ . \nn
\eeqa
This expansion of  fermions is in straight analogy with the expansion \eqref{eq:TKM}. In fact,  introducing $b^i_m(\varphi)$,
we simply  assume that the usual string modes of fermions have an expansion in the $\varphi-$direction.

\medskip
As the two-torus has two non-contractible loops, we have four kinds of sectors: (R,R), (NS,R), (R,NS), (NS,NS), corresponding to periodic (Ramond $-$R) or anti-periodic (Neveu-Schwarz $-$NS) boundary conditions. The summation index belongs to $ \mathbb Z$ (resp.  to $\mathbb Z + \nicefrac 12$) for R (resp. NS)  fermions.
 It is straightforward to verify that the relation
\beqa
\label{eq:laur}
b^i_{mp}= \oint \frac{ \text{d} z\; z^{m}}{2 \ii \pi z}\int\limits_{0}^{2\pi} \frac{\text{d} \varphi}{2 \pi} e^{\ii p \varphi} H^i(z,\varphi) \ 
\eeqa
is satisfied. We postulate the quantization relations
\beqa
\label{eq:quantS1S1}
\{b^i_{mp},b^j_{nq}\}=\delta^{ij} \delta_{m+n}\delta_{p+q} \ ,
\eeqa
and the reality condition
\beqa
\label{eq:real}
(b^i_{mp})^\dag = b^i_{-m-p} \ .
\eeqa
 To define annihilation and creation operators we introduce the following order relation:
\beqa
\label{eq:ord}
(m,p)>0 \Longleftrightarrow \Big( m>0 \ \ \text{or} \ \ (m=0 \ \ \text{and} \ \ p>0)\Big) \ . 
\eeqa
The vacuum is thus defined by
\beqa
\label{eq:VacT}
\begin{array}{ll}
\begin{array}{rlll}
b^i_{mp} \big|0\big>&=&0\ , &  m>0 \ , \ \ \forall  p \in \mathbb Z\; (+\frac 12)   \\[0.25cm]
b^i_{0p}  \big|0\big>&=&0\ , &  p>0\ .
\end{array}&  \
\end{array}
\eeqa
 We observe that in the (R,R) sector, due to the fact that $b^i_{00}, i=1,\cdots, d$ generate a Clifford algebra, the vacuum lies in the $2^{\big[\nicefrac d 2\big]}-$dimensional spinor representation of $\mathfrak{so}(d)$ (here $[x]$ denotes the integer part of $x$).

\medskip
In order to proceed further, we define the normal ordering prescription:
\beqa
\label{eq:NO}
 {}_\circ \hskip -.17truecm {}^\circ \hskip .1 truecm b^i_{mp}  b^{j}_{nq}\no = \left\{
 \begin{array}{cll}
 -  b^{j}_{n q} b^i_{mp} &\text{if~} m>0&\forall p \in \mathbb Z \ (+\frac12)\\
 \frac12( b^i_{mp}b^{j}_{nq}-b^{j}_{n q} b^i_{mp}) &\text{if~} m=0&\forall p  \in \mathbb Z \ (+\frac12) \\
  b^i_{mp}   b^{j}_{nq}&\text{if~} m<0&\forall p  \in  \mathbb Z \ (+\frac12)\\
\end{array}\
\right.
\eeqa
This normal ordering prescription   is  standard \cite{go, dms}.
However,  it turns out that \eqref{eq:NO}  is not sufficient for our construction.  As we will see, this is a consequence
  of infinite sums associated to the second loop of the two-torus (the $\varphi-$direction).

Having introduced fermions on the two-torus, we now construct a generalised Kac-Moody  algebra  and a generalised Virasoro algebra associated to
$\mathbb S^1 \times \mathbb S^1$. As the construction involves bilinear fermions, we have to proceed carefully {\bb and, in particular, we have to regularise infinite sums}. To this extent, we introduce a cut-off and define `$\epsilon-$regularised' fermions ($\epsilon>0$): 
\beqa
\label{eq:Heps}
H_\epsilon^i (z,\varphi) =
\sum  \limits_{m\in \mathbb Z (+\frac12)} \sum  \limits_{p\in \mathbb Z (+\frac12)} b_{mp}^i z^{-m}  e^{- \ii p\varphi}e^{-\epsilon(|p| -\frac12)} \ ,
\eeqa
the $\epsilon-$term  being necessary to regularise infinite sums in the $\varphi-$directions.
Using the normal ordering prescription \eqref{eq:NO}  and the quantization relation \eqref{eq:quantS1S1},
it follows from  application of the Wick theorem~\cite{wick} that
\beqa
\label{eq:Heps2}
H_\epsilon^i(z,\theta) H_\epsilon^j(w,\varphi)= \no H_\epsilon^i(z,\theta) H_\epsilon^j(w,\varphi)\no + \delta^{ij}\Delta(z,w)
\delta_\epsilon(\theta -\varphi) \ , \ \ |z| >|w|
\eeqa
where 
\beqa
\label{eq:deleps}
\delta_\epsilon(\theta -\varphi) = \sum \limits_{m \in \mathbb  Z  (+ \frac 12)} e^{-\ii m (\theta-\varphi)} e^{-2 \epsilon(|m|-\frac12)}\ ,
\eeqa
and
\beqa
\label{eq:Del}
\Delta(z,w)= \left\{
\begin{array}{ll}
\sum\limits_{p=\frac 12 }^{+\infty} \left(\frac{w}{z}\right)^p = \frac{\sqrt{zw}}{z-w}&  \text{NS}\\
\sum\limits_{p=1}^{+\infty} \left(\frac{w}{z}\right)^p + \frac 12= \frac 12\frac{z+w}{z-w}&  \text{R}
\end{array}
\right. \ \  \ \ |z| >|w|
\eeqa
(here R or NS stands for the behaviour of fermions in the $z-$direction).
We observe that for $\epsilon \to 0$, on the one hand $H_\epsilon^i (z,\theta) \to H^i (z,\theta)$, while on the other hand the function $\delta_\epsilon$ reduces to the usual $\delta-$distribution.

However, as the generators of
the Kac-Moody and the Virasoro algebras on the two-torus are bilinear in the fermions (see below), the Wick theorem involves a double contraction
and in particular  terms of the type $\delta_\epsilon(\varphi-\theta)^2$.   This term is perfectly defined, while  its limit for $\epsilon \to 0$ is not defined.  In order to circumvent this difficulty and obtain a well-defined limit, we {\bb define a regularisation procedure.}   This {\bb regularisation} is done in two steps. In the first  step, we substitute $\delta_\epsilon(\theta-\varphi)^2$ by  $\delta_{\epsilon_1} (\theta-\varphi)\delta_{\epsilon_2}(\theta-\varphi)$ and take the limit of $\epsilon_1\to 0, \epsilon_2\to 0$ independently.  In other words ,
\beqa
\lim_{\epsilon_1 \to 0} \delta_{\epsilon_1} (\theta-\varphi)\delta_{\epsilon_2}(\theta-\varphi)=  \delta (\theta-\varphi)\delta_{\epsilon_2}(\theta-\varphi)=  \delta (\theta-\varphi)\delta_{\epsilon_2}(0) \ . \nn
\eeqa
In a second step, the divergent limit $\epsilon_2\to 0$  for $\delta_{\epsilon_2}(0)$ is regularised using the $\zeta-$regularisation (with $\zeta$ being the Riemann $\zeta-$function): 
\beqa
\label{eq:zet}
\delta_{\text{reg.}}(0)=
\left\{
\begin{array}{llll}
2 \sum\limits_{p=0}^{+\infty} e^{-2\epsilon_2(p -\frac12)}-1 &\to& 2\zeta(0,-\frac 12)-1=1&\text{R}\\
2 \sum\limits_{p=0}^{+\infty} e^{-2\epsilon_2 p} &\to& 2\zeta(0,0)=1&\text{NR}
\end{array}\right.
\eeqa
where R, NS stand for the behaviour of the fermions in the $\varphi-$direction, {\it i.e.}, when $p\in \mathbb Z$ or $p\in \mathbb Z +
\nicefrac1  2$.  {\bb This means that, within this regularisation procedure, we obtain
  \beqa
 \lim _{\epsilon \to 0} \delta^2_\epsilon(\theta-\varphi)_{\text{reg}} = \delta(\theta-\varphi) \ . \nn
  \eeqa\\
} 
We now introduce the generators of the Kac-Moody and the Virasoro algebras associated to $\mathbb S^1 \times \mathbb S^1$. Let $\g$ be
a compact Lie algebra and let ${\cal D}$ be a real unitary $d-$dimensional representation  (the case where the representation is not real can also be considered, as shown in \cite{go}, where all results proved below can be extended without major complications to this case).  Denote
the generators of $\g$ in the representation ${\cal D}$ by  the real antisymmetric matrices $M_a, a =1\,\cdots, \dim \g$. We thus have
\beqa
\big[M_a, M_b\big] = f_{ab}{}^c M_ c \ , \ \  M_a^\ast = M_a\ , \ \ M_a^t=-M_a \ . \nn
\eeqa
Define now the operators
\beqa
\label{eq:KM-VIr}
T_\epsilon^a(z,\theta)&=&\frac \ii 2 M^a_{ij}\noo H_\epsilon^i(z,\theta) H_\epsilon^j(z,\theta) \noo\nn\\
L_\epsilon(z,\theta)&=& \frac 12 \sum \limits_{i=1}^{d} \noo z \frac{\text{d} H_\epsilon^i(z,\theta)}{\text{d} z} H_\epsilon^i(z,\theta)\noo + \lambda d
\eeqa
where $\lambda=0$ (resp. $\lambda=\nicefrac{1}{16}$) for fermions with NS (resp. R) boundary conditions in the $z-$direction.
These operators are a direct generalisation  of the Virasoro and Kac-Moody algebras bilinear in fermions
obtained in fermionic strings \cite{GSW1, go}.
We have introduced the  regulator $\epsilon$, {\bb as described above}.
{\bb In particular, if we compute  $T_\epsilon T_\epsilon$, we get
  \beqa
  T_\epsilon(z,\theta) T_\epsilon(w,\varphi)&=&\no T_\epsilon(z,\theta) T_\epsilon(w,\varphi)\no
 -\frac 1 4 M_{ij}^a M_{k\ell}^b\Delta(z,w) \delta_\epsilon(z-w)\times
\nn\\
 &&
 \Big( \delta^{jk}\;\no H^i(z) H^\ell(w) + \delta^{i\ell} \;\no H^j(z) H^k(w)-\delta^{ik} \; \no H^j(z) H^\ell(w)\no \nn\\
&&
-\delta^{j\ell}\; \no H^i(z) H^k(w)\no\Big) + \frac12 C_M \delta^{ab} \Delta^2(z,w) \delta^2_\epsilon(\theta-\varphi)\ ,  |z| >|w| \ .\nn
 \eeqa
If we now take the limit $\epsilon\to0$, all terms have a finite limit except the last one, obtained from the double contraction. 
Thus, in order to take the limit $\epsilon \to 0$, we have to regularise the infinite sums using the prescription described before, {\it i.e.}, by means of the $\zeta-$regularisation \eqref{eq:zet}.}
{\bb Proceeding similarly for all other terms we obtain, after having extracted the divergent part (omitting the regular part)}
\beqa
\label{eq:KMVir}
T^a(z,\theta) T^b(w,\varphi)&=&
 -\frac 1 4 M_{ij}^a M_{k\ell}^b\Delta(z,w) \delta(z-w)
\Big(\delta^{jk}\;\no H^i(z) H^\ell(w)\no \nn\\
&&
 + \delta^{i\ell} \;\no H^j(z) H^k(w)-\delta^{ik} \; \no H^j(z) H^\ell(w)\no \nn\\
&&
-\delta^{j\ell}\; \no H^i(z) H^k(w)\no\Big) + \frac12 C_M \delta^{ab} \Delta^2(z,w)\delta(\theta-\varphi)\ ,  |z| >|w| \ .\nn\\
 L(z,\theta) L(w,\varphi)&=&
 d\Big( \frac14 z\partial_z \Delta(z,w)  w\partial_w \Delta(z,w) -\frac14 \Delta(z,w) zw \partial_z \partial_w\Delta(z,w)\\
&& \hskip .2truecm - 2\lambda w\partial_w \Delta(z,w)\Big)\delta(\theta-\varphi)
 +2 L(w,\varphi) w \partial_w \Delta(z,w)\delta(\theta-\varphi) \nn\\
&&\hskip .2truecm+ \partial_w L(w,\varphi)\Delta(z,w)\delta(\theta-\varphi)
\ , \ \ |z| > |w| \ , \nn\\
L(z,\theta) T^a(w,\varphi)&=& T^a(w,\varphi) w \partial_w \Delta(z,w) \delta(\theta-\varphi) +
w \partial_w T^a(w,\varphi) \Delta(z,w) \delta(\theta-\varphi) \ , \nn\\
\ \ |z| > |w| \nn \ .
\eeqa
with
\beqa
\label{eq:Tr}
\text{Tr}(M_a M_b) = -\delta_{ab} C_M\ \  ,\ \  C_M > 0 \ .
\eeqa
Extracting the modes
\beqa
\label{eq:TaLn}
T^a_{mp}=  \oint \frac{ \text{d} z z^{m}}{2 \ii \pi z}\int\limits_{0}^{2\pi} \frac{\text{d} \varphi}{2 \pi} e^{\ii p \varphi} T^a(z,\varphi) \ , \ \ 
L_{mp}=\oint \frac{ \text{d} z z^{m}}{2 \ii \pi z}\int\limits_{0}^{2\pi} \frac{\text{d} \varphi}{ 2 \pi} e^{\ii p \varphi} L(z,\varphi) \ 
\eeqa
standard techniques of Conformal Field Theory, {\it i.e.},\   integration
in the complex $z-$ and $w-$planes with adapted integration contours (see {\it e.g.} \cite{go, dms}) 
lead to (the  $\varphi$ and  $\theta$ integrations are trivial)
\beqa
\label{eq:TLS1S1}
\big[T^a_{mp}, T^b_{nq}\big]&=&if^{ab}{}_c T^c_{m+n p+q} + \frac 12 C_M m \delta^{ab}\delta_{m+n}\delta_{p+q} \nn\\
\big[L_{mp},L_{nq}\big]&=& (m-n) L_{m+n p+q} + \frac {d/2}{12}m(m^2-1) \delta_{m+n}\delta_{n+q}\\
\big[L_{mp}, T^a_{nq}\big]&=&-p T^a_{m+np+q} \nn
\eeqa
It follows that \eqref{eq:TLS1S1} is a fermionic realisation of the semi-direct product  of the Kac-Moody  and Virasoro algebras of the two-torus
$\hat{\g}(\mathbb S^1 \times \mathbb S^1) \rtimes
\text{Vir}(\mathbb S^1 \times \mathbb S^1)$   with central charges  equal to $c=d/2, k=C_M/2$ respectively.
A direct consequence is  that the full representation obtained acting on the vacuum \eqref{eq:VacT} is unitary. 
This algebra obviously
contains  the  semi-direct product of the usual Virasoro and Kac-Moody algebras as a subalgebra.  As mentioned before, the fermionic realisation is dissymmetric in both directions of the two-torus. This is a  consequence of the ordering relation \eqref{eq:ord}, remaining in accordance with our results of \cite{rmm}, where it was proved that unitarity of representations of the Kac-Moody algebra of the two-torus implies  that only  one central charge is non-vanishing.

\medskip
 At a first glance, it may appear that our regularisation
 procedure \eqref{eq:zet} looks awkward.
 {\bb First of all, it seems  that the use of $\zeta-$regularisation can be dangerous in another (but related) context.  For instance,  suppose we have a  system of $N$ pairs of fermions with large $N$:
   $H^i_\epsilon(z,\varphi), i=1,\cdots ,2N$. Intuitively it seems that the central charge of the Virasoro algebra, which is $c=N$ (for finite $N$), becomes $1$ after regularisation, whenever $N$ goes to infinity. Actually, this is not the case. Indeed, considering more carefully our regularisation procedure, the regularisation is made upon the $\varphi-$direction ({\it i.e.}, the modes of $H^i(z,\varphi)$,
   see Eq.[\ref{eq:Heps}])  {\it and not} upon the summation over the number of fermions ({\it i.e.}, over the index $i$ associated to the `flavour' of fermions), as we do not have an analogue of \eqref{eq:Heps} related to the summation over the index $i$ when $N\to \infty$. Thus, in particular, if one takes the limit $d= 2N \to \infty$, we observe that the central charge in $ L L$ diverges and cannot be regularised (see \eqref{eq:KMVir}), as expected. Under the same limit, the central charge associated to the Kac-Moody algebra which involves the trace of $2N\times 2N$ matrices (see Eq.[\ref{eq:Tr}]) also diverges, as it should be.  A similar phenomenon arises when considering the zero-modes of {\bb fermions}. In the (R,R)
   sector, the zero modes are associated to the fermion operators $b_{00}^i, i=1,\cdots, {\bb 2}N$ (see \eqref{eq:VacT}). When $N$ is finite, the  zero modes lead to the spinor representation of
   $\mathfrak{so}(2M)$, with  a finite number of states for one-particle states, two-particles states, {\it etc}, whereas it is
   not the case when $d=2 N\to \infty$, since at each level we have an infinite number of states. Of course, in this context the $\zeta-$function regularisation will {\it never solve} the problem, as we have an infinite number of states. This in particular means
   that our procedure {\it cannot} be extended for an infinite number of fermions $H^i(z,\varphi)$ or an infinite number of flavours. To conclude this observation,
   we still have to emphasise an important distinction. We have a finite number of regulated-fermions $H_\epsilon^i(z,\varphi)$ with $i=1,\cdots, d=2N$, thus for fermions in the (R,R) sector the situation described above never happens. When we expand the modes of   $H_\epsilon^i$ in the $\varphi-$direction we use \eqref{eq:Heps} and obtain an infinite number of modes.  Since now we have an infinite number of modes, the divergences associated to the expansion in the $\varphi-$direction are regulated by the $\zeta-$regularisation.

{\bb   We now analyse the Hilbert space more carefully.  To this extent, we slightly modify the definition of the vacuum for the (R,R) and (R, NS) sectors, which is defined as follows:
   \beqa
   b^i_{mp} \big|0\big>&=&0 \ , i\ =1 ,\cdots, 2N \ ,  \ \ \forall p \in \mathbb Z + (\frac 12),\ \  m>0 \nn\\
   a^i_p \big|0\big>&=&0 \ ,  i=1,\cdots, N\ , \ \ \forall p \in \mathbb Z + (\frac 12) \ , \nn
 \eeqa
 with $a^i_p=\nicefrac 1{\sqrt 2} (b^i_p + i b^{i+N}_p)$; the vacuum being unchanged for the (NS,R) and (NS,SN) sectors.
The Hilbert space is constructed with   infinitely many fermions corresponding to the mode expansion $b_{mp}^i$ of the fermions
         on the two-torus $H^i(z,\varphi), i=1,\cdots, d$ which satisfy the anticommutation relation \eqref{eq:quantS1S1} and the reality condition \eqref{eq:real}. It is thus straightforward to construct the corresponding Fock space in terms of the fermionic oscillators  $b_{mp}^i$. Stated differently, if we consider the fermions operators  $H^i(z,\varphi)$ and $H_m^i(\varphi)$ some care must be taken as we now see.
 If we define:
 \beqa
 \big|i,m,\varphi\big> &=& H^i_{-m}(\varphi) \big|0\big>  \ , m >0 \ , i=1, \cdots, 2N \nn\\
 \big|i,\varphi\big> &=& \frac 1 {\sqrt{2}} \big(H_0^i(\varphi) -i H_0^{i+N}(\varphi)\big) \big|0\big> \ , i=1, \cdots, N \nn
 \eeqa
 and it is easy to see that we have
 $
\big<i,m,\varphi\big|j,n, \theta\big>=\delta^{ij} \delta_{mn} \delta(\varphi-\theta) \ \ \text{and}\ \ \big<i,\varphi\big|j,\theta\big>=\delta^{ij}  \delta(\varphi-\theta). 
$
Similarly, if  we consider $\big|i,z,\varphi\big> = H^i(z,\varphi)\big|0\big>$, we have
      $\big<i,z,\varphi\big|j,w,\theta)\big>= \delta^{ij} \Delta(z,w) \delta(\varphi-\theta)$.
This construction extends to the $p-$particle subspace of the Hilbert space, but
since the operators $H^i(z,\varphi)$ and $H_m^i(\varphi)$ involve an infinite number of fermions, as usual in QFT, the normal ordering prescription and the regularisation procedure have to be applied
(in particular for the latter, when we have operators as the same points, as  happens for instance in \eqref{eq:KM-VIr}). Thus,  the normal ordering prescription (but with possibly the additional
$\zeta-$regularisation) must be considered when computing $K-$points functions. 
 \footnote{\bb We would like to thank the referee for pointing out these subtle points related to the structure of Hilbert space.}

  To conclude this analysis, let us mention that the infinitely many fermions corresponding to the modes expansion of the fermions
  on the two-torus $H^i(z,\varphi), i=1,\cdots, d$ enable us to construct in a  straightforward way the Hilbert space (some  caution is required only when we consider fermions operators with an infinite number of fermions modes $b_{mp}^i$, as for instance $H^i_m(\varphi)$ or $H^i(z,\varphi)$).
         In this construction there is no need to proceed to  any normal ordering or regularisation prescription, since the action of the fermion modes on the Fock space is well defined.
          In other words,  a normal ordering and  regularisation prescription are needed whenever one tries to define a Kac-Moody and a Virasoro algebra associated to the two-torus, as we have operators bilinear in $H^i(z,\varphi)$, in particular for the computation of the central charge. Having regulated the algebra, it turns out the Fock space associated to the fermions is naturally a unitary representation of the Kac-Moody and Virasoro algebras of the two-torus. This means that  we have, prior to a fermionic realisation of our algebras as in \eqref{eq:KM-VIr},
a unitary representation.  In this context, it should be mentioned that in \cite{KR},  
(unitary) representations of certain infinite dimensional Lie algebras (occurring in the physical literature as $W_{1+\infty}-$algebras) are obtained using more sophisticated techniques  based on quasi-finite highest weight modules, but that this heavy machinery is not required in our approach.}}

{\bb  The reason for which we have introduced a  $\zeta-$regularisation beyond the standard normal ordering prescription  is simply due to the fact that the mode expansion of our operators involves a summation over $\mathbb Z^2$. In a different approach, initiated in \cite{MRT}, this double summation has been avoided. Basically, the main difference  is related to the presentation of the algebra.  Actually, the algebra used in \cite{MRT} and called the toroidal algebra, is presented through a centrally extended loop algebra of the affine Lie algebra $\hat \g$ (associated
  to the simple Lie algebra $\g$), whilst  we are concerned  with a centrally extended algebra of the double loop algebra of the simple Lie algebra $\g$. Thus, in \cite{MRT} the presentation of the algebra is given through the roots of $\hat\g$, whereas in our approach the algebra is defined though the roots of  $\g$. The  approach of \cite{MRT} has been followed in  \cite{jk}, where a fermionic representation of certain representations of the simple Lie algebras $\mathfrak a_n,  \mathfrak b_n, \mathfrak c_n,
  \mathfrak d_n$ has been obtained.
  Thus, a Clifford algebra (or  Neveu-Schwarz-fermionic field) associated to  any roots $\alpha$ of $\hat\g$  and to $m\in \mathbb Z + \frac 12$ 
  $f_\alpha(m)$  satisfying 
  \beqa
  \label{eq:cliff}
\{f_\alpha(m),f^\dag_\beta(n)\}= \delta_{m+n} \alpha\cdot \beta \ , 
  \eeqa
  is introduced. These operators enable to construct explicitly  some  fermionic representation of  the toroidal algebra.
This construction is very close to our construction with (at least) two differences.
Firstly, since the
expansion involves only one number (associated to the modes of the circle), standard techniques of Conformal Field Theory can be used
without the  $\zeta$-regularisation. Secondly, as it is well known, the determinant of the Cartan matrix of affine Lie algebras equals zero
(which is not the case for a simple Lie algebra),
meaning that the root system contains imaginary (or null roots), generally denoted $n \delta, n \in \mathbb Z\setminus\{0\}$ and such that
$\delta\cdot \delta=0$. Thus the Clifford algebra \eqref{eq:cliff} is degenerate and the representation space contains zero-norm states,  implying that the representations obtained in \cite{jk} are non unitary. An interesting question that deserves further investigation is the {\bb detailed comparison of the approach presented in this work and that proposed in \cite{jk}}, and, in particular, how to deal with zero-norm states.}

   {\bb In a different context}, it should be mentioned that the appearance of unwanted terms like $\delta^2$ is not
  an
uncommon phenomenon, always requiring an appropriate regularisation procedure consistent with their physical interpretation. In this context, terms of this type appear for example in standard Quantum Field Theory, specifically in connection with the conservation of momentum when computing cross sections, where these terms are managed consistently by other techniques. In the case that occupies us, we observe that the coefficient $\ell = \nicefrac 1{16}$ in Eq.[\ref{eq:KM-VIr}] (in the Ramond $z-$sector) is the same as in fermionic strings, and comes from a regularised term in the double contraction. Besides, the $\zeta-$regularisation ensures that we obtain the correct central extensions for $\hat {\g}$ and Vir as subalgebras of their analogue on the two-torus, hence proving to be a tool that is consistent with the underlying algebraic extension formalism. In this frame, we recall that the use of the $\zeta-$function in regularisation problems has already shown its utility in string theory, where it has been applied successfully to regularise the quantum numbers of the vacuum \cite{rasa}.

\medskip
We now turn to the fermions on the two-sphere $\mathbb S^2$, which is a much more involved case. Again, with the purpose in mind to use  techniques of Conformal Field Theory and Wick rotations, we substitute $(z,u)$ to the usual
 spherical coordinates $(\varphi,u)$
  (with $u=\cos \theta$). As for the two-torus, two kinds of fermions can be considered (R or NS), but only in the $z$ (or $\varphi$) direction. Since we also have to regularise infinite sums we introduce, as
  for $\mathbb S^1 \times \mathbb S^1$, $\epsilon-$regularised fermions. In the R-sector the $\epsilon-$regularised fermions are defined by
\beqa
H_\epsilon^i(z,u)=
 \sum\limits_{m\in \mathbb Z}  b_{\epsilon,m}^i(u) z^{-m}\ , \ \ b_{\epsilon, m}^i(u)  = \sum\limits_{\ell=|m|}^{+\infty} b^i_{\ell,m}  Q_{\epsilon,\ell m}(u)\nn
\eeqa
where $Q_{\epsilon, \ell m}$   ($\ell \in \mathbb N, -\ell\le m \le \ell$) are the $\epsilon-$`regularised' associated Legendre functions
\beqa
\label{eq:Leg}
Q_{\epsilon, \ell m}(u) = e^{-\epsilon a_m}\frac{(-1)^{\ell +m}}{2^\ell \ell !} \sqrt{2\ell +1}\sqrt{\frac{(\ell-m)!}{(\ell+m)!}} (1-u e^{-\epsilon})^{m/2}
\frac{d^{\ell +m}}{du^{\ell +m}} \big[1-u^2 e^{-2\epsilon})^\ell\big]\ ,\nn\\
\eeqa
 with $a_m$ a constant to be determined.
The corresponding $\delta^m_\epsilon$ functions (one for each value of $m$) are then defined by
\beqa
\delta^m_\epsilon(u-v) = \sum_{\ell \ge |m|} Q_{\epsilon, \ell m}(u)  Q_{\epsilon, \ell m}(v) \nn
\eeqa
When $\epsilon\to 0$  on the one hand $H^i_\epsilon  (z, u) \to H^i (z, u)$ (the non-regularised fermions)
and, on the other hand, the functions $\delta_\epsilon^m$ reduce to the usual $\delta-$distribution \eqref{eq:delS2R}.
 Extracting the modes we have, using the basis \eqref{eq:Bn} 
\beqa
\label{eq:laurR}
b^i_{mn}= \oint \frac{ \text{d} z\; z^{m}}{2 \ii \pi z}\int\limits_{-1}^{1} \frac{\text{d} u}{2} Q_{\ell m}(u) H^i(z,u) \ .
\eeqa

The NS fermions, {\it i.e.}, anti-periodic in the $z-$direction, are defined by 
\beqa
H_\epsilon^i(z,u) &=& \frac 1 {\sqrt 2}\sum \limits_{m\in \mathbb Z + \frac12} \Bigg(\sum \limits_{\ell=|m|}^{+\infty} \sum \limits_{\eta=\pm}
b^{i \eta}_{\ell, m} Q^{\epsilon,(|m-\frac \eta 2|, |m+\frac \eta 2|)}_{\ell-|m|}(u) \Bigg) z^{-m}\nn\\
&=& \frac 1 {\sqrt 2}\sum\limits_{m\in \mathbb Z +\frac12}
 b_{\epsilon,m}^i(u) z^{-m}\ , \nn
\eeqa
where
\beqa
\label{eq:Jac}
Q^{\epsilon,(|m-\frac \eta 2|, |m+\frac \eta 2|)}_{\ell-|m|}(u)&=& \frac{\sqrt 2}{2^\ell}
                         \frac{ (-1)^{\ell -|m|}}{( \ell -\frac12)!}
\sqrt{\frac{(\ell +|m|)!}{(\ell-|m|)!}}                 
\frac {e^{-\epsilon a_m}}  {\sqrt{(1-u \;e^{-\epsilon})^{|m-\frac \eta 2|}(1+ u \;e^{-\epsilon} )^{|m+\frac \eta 2|}}} \nn\\
&&\frac{\text{d}^{\ell -|m|}}{\text{d} u^{\ell -|m|}}\big[
(1-u^2 \;e^{-2 \epsilon})^{\ell -|m|}(1-u \;e^{-\epsilon})^{|m-\frac \eta 2|}(1+u \;e^{-\epsilon})^{|m+\frac \eta 2|}\big] \ , \nn\\
\eeqa
with again $a_m$ a constant to be determined. 
Note that if we  permute $(m,\eta)$ with $(-m,-\eta)$, we get the equality
$
Q^{\epsilon,(|m-\frac \eta 2|, |m+\frac \eta 2|)}_{\ell-|m|}(u)=
Q^{\epsilon,(|-m+\frac \eta 2|, |-m-\frac \eta 2|)}_{\ell-|-m|}(u).
$ 
The corresponding $\delta_\epsilon^{\eta,m}$ functions are given by
\beqa
\delta_\epsilon^{\eta,m}(u-v) = \sum \limits_{\ell \ge |m|}  Q^{\epsilon,(|m-\frac \eta 2|, |m+\frac \eta 2|)}_{\ell-|m|}(u)Q^{\epsilon,(|m-\frac \eta 2|, |m+\frac \eta 2|)}_{\ell-|m|}(v)\eeqa
It is routine to verify that for $\epsilon\to 0$,
$\Big((1-u)^{|m-\frac \eta 2|}(1+ u)^{|m+\frac \eta 2|}\Big)^{-1/2}$ $ Q^{\epsilon=0,(|m-\frac \eta 2|, |m+\frac \eta 2|}_{\ell-|m|}(u)$  are  Jacobi polynomials
\cite{Abri,nu,be}. In \cite{Abri}  it was shown that the solution of Dirac equation on $\mathbb S^2$ is expressed in terms of  Jacobi polynomials.  For $\epsilon =0$,
 and for any value of $m$ and  $\eta=\pm$,  the set ${\cal B}_{\eta,m}= \big\{Q^{\epsilon=0,(|m-\frac \eta 2|, |m+\frac \eta 2|)}_{\ell-|m|}, \ell\ge |m|\big\}$ forms a Hilbert orthonormal basis (with ${\cal B}_{m,\eta}= {\cal B}_{-m,-\eta}$).
Further,  for $\epsilon\to 0$ we have that 
$H^i_\epsilon  (z, u) \to H^i (z, u)$ (the non-regularised fermions), while on the other hand the functions $\delta_\epsilon^{\eta,m}$ reduce to the usual $\delta-$distribution:
\beqa
\delta(u-v)=  \sum \limits_{\ell \ge |m|}  Q^{\epsilon=0,(|m-\frac \eta 2|, |m+\frac \eta 2|)}_{\ell-|m|}(u)Q^{\epsilon=0,(|m-\frac \eta 2|, |m+\frac \eta 2|)}_{\ell-|m|}(v)
\eeqa
 We thus get the integral expression
\beqa
\label{eq:laurNS}
b^{i,\eta}_{\ell m}= \sqrt 2 \oint \frac{ \text{d} z\; z^{m}}{2 \ii \pi z}\int\limits_{-1}^{1} \frac{\text{d} u}{2 }   Q^{\epsilon=0,(|m-\frac \eta 2|, |m+\frac \eta 2|)}_{\ell-|m|}(u)H^i(z,\varphi) \ .
\eeqa

 For both the R and NS sectors the  quantisation relations  are taken to be
\beqa
\label{eq:quantS2}
\begin{array}{ll}
\{b_{\ell_1,m_1}^{i_1},b_{\ell_2,m_2}^{i_2}\} = (-)^{m_1} \delta^{i_1 i_2} \delta_{\ell_1\ell_2} \delta_{m_1+m_2} \ , & \text{R}\ ,\\[0.2cm]
\{b_{\ell_1,m_1}^{\eta_1,i_1},b_{\ell_2,m_2}^{\eta_2,i_2}\} =  \delta^{i_1 i_2} \delta_{\ell_1\ell_2} \delta_{m_1+m_2} \delta_{\eta_1 + \eta_2}\ ,
& \text{NS}\ ,
\end{array}
\eeqa
and the reality condition
\beqa
\begin{array}{ll}
 (b_{\ell,m}^i)^\dag = (-1)^m b_{\ell,-m}^i,&\text{R}\\
 (b_{\ell,m}^{\eta,i})^\dag = b_{\ell,-m}^{-\eta,i},&\text{NS} \ .
 \end{array} \nn
\eeqa
In the case of R (resp. NS) fermions, we define for $m>0$  the fermionic annihilation/creation operators $f^i_{\ell,m}=b^i_{\ell,m}, f^{i \dag}_{\ell,m}=(-)^m b^i_{\ell,-m}$
(resp, $f^{i,\eta}_{\ell,m}=b^{i,\eta}_{\ell,m}, f^{i,\eta \dag}_{\ell,m}=b^{i,-\eta}_{\ell,-m}$)  associated to the fermionic fields $H^i(z,u)$.  Note also that the generators $f^i_{\ell,0} = f^{i\dag}_{\ell,0}$ generate a Clifford algebra. The vaccuum state, $\big|0\big>$, as usual, is annihilated
by all annihilation operators. We next define  the normal ordering prescription for R-fermions:
\beqa
\label{eq:NOS21}
\no b_{\ell,m}^i b^j_{\ell',m'}\no= \left\{
\begin{array}{cl}
-b^j_{\ell',m'} b^i_{\ell,m}&m>0\ , \forall \ell \ge m\\
\frac 12( b^i_{\ell,m}b^j_{\ell',m'} -b^j_{\ell',m'} b^i_{\ell,m})&m=0\ , \forall \ell \in \mathbb N\\
 b^i_{\ell,m}b^j_{\ell',m'}&m<0\ , \forall \ell \ge -m
 \end{array}
\right.
\eeqa
and for the NS fermions
\beqa
\label{eq:NOS22}
\no b^{i, \eta}_{\ell,m} b^{j,\eta'}_{\ell',m'} \no=\left\{
\begin{array}{ll}
-b^{j,\eta'}_{\ell',m'} b^{i, \eta}_{\ell,m}&m>0\ ,\  \forall \ell>m\phantom{-}\  ,\  \eta=\pm\\
\phantom{-} b^{i, \eta}_{\ell,m}b^{j,\eta'}_{\ell',m'}&m<0\ ,\  \forall \ell>-m\ ,\  \eta=\pm
 \end{array}
 \right.
\eeqa
Using the normal ordering prescription \eqref{eq:NOS21} and \eqref{eq:NOS22} and the quantisations relations \eqref{eq:quantS2},  application of the Wick theorem leads to
\beqa
H_\epsilon^i(z,u) H_\epsilon^j(w,v) = \no H_\epsilon^i(z,u) H_\epsilon^j(w,v) \no + \delta^{ij} \Delta_\epsilon(z,w;u,v) \ , \ \ |z|> |w| \nn
\eeqa
where
\beqa
 \Delta_\epsilon(z,w;u,v) =\left\{
 \begin{array}{ll}
 \hskip .4truecm \sum \limits_{m>0}  \delta_\epsilon^m(u-v) \Big(\frac w z \Big)^m + \frac 1 2  \delta_\epsilon^0(u-v)& \text{R}\\
 \frac 12\sum \limits_{m\ge1/2} \sum \limits_{\eta=\pm} \delta^{\eta,m}_\epsilon(u-v) \Big(\frac w z \Big)^m& \text{NS}
 \end{array}
 \right. \ \ \ \ |z| > |w|
\eeqa
As in the case of the two-torus, all expressions are perfectly defined in the limit when $\epsilon\to 0$. In particular we have in this limit
$\Delta_\epsilon(z,w;u,v)\to \Delta(z,w) \delta(u-v)$ with $\Delta(z,w)$ defined in \eqref{eq:Del}. However,
as happened for the two-torus, 
since the generators of the Kac-Moody and the Virasoro algebras on the two-sphere are bilinear in the fermions, the Wick theorem involves a double contraction, and in particular, terms like $\Delta_\epsilon(z,w,u,v)^2$.
Albeit this term is perfectly defined, it is not for the limit $\epsilon \to 0$.
As before, we consider a {\bb  regularisation} prescription in two steps to surmount this difficulty. In the first step, we substitute $\Delta_\epsilon(z,w,u,v)^2$ by  $\Delta_{\epsilon_1} (z,w,u,v)\Delta_{\epsilon_2}(z,w,u,v)$ and take the limit $\epsilon_1\to 0, \epsilon_2\to 0$ independently.  In a second step, we regularise all $\delta_\epsilon(0)$ given by
\beqa
\begin{array}{llll}
\delta^m_\epsilon(0) &=& \sum \limits_{\ell\ge |m|} \Big(Q_{\epsilon,\ell m}(0)\Big)^2&\text{R}\\
\delta^{\eta, m}_\epsilon(0) &=& \sum \limits_{\ell\ge |m|} \Big(Q^{\epsilon,(|m-\frac \eta 2|, |m+\frac \eta 2|)}_{\ell-|m|}(0)\Big)^2&\text{NS}
\end{array}
\eeqa
Using the Stirling formula and the asymptotic expressions for large values of $\ell$
\beqa
Q^{ 2}_{\epsilon,\ell m}(0)  &\sim& \left\{\begin{array}{cl} \frac 4 \pi  e^{-2\epsilon(\ell + |m|+a_m)}& \ell +m \ \text{even}\\
                                                          0& \ell +m \ \text{odd}\end{array}\right.\nn\\
\hskip 4.truecm \text{~and} \\
(Q^{\epsilon,(|m-\frac \eta 2|, |m+\frac \eta 2|)}_{\ell-|m|}(0))^{ 2}&\sim&
\frac 2 {\pi \ell}   e^{-2\epsilon(\ell - |m|+a_m)}\nn
\eeqa
we obtain
\beqa
\delta^m_\epsilon(0)&=& C_m + \sum_{\begin{array}{c} \ell \ge 0\\
\ell + m \ even \end{array}} \frac 4 \pi {  e^{-2\epsilon( \ell + |m| +a_m)}}\ \ \text{R} \nn \\
\delta^{\eta,m}_\epsilon(0) &=& 
C_{\eta,m} + \sum_{\ell >0} \frac 2 {\pi \ell} e^{ -2 \epsilon(\ell -|m| +a_m) } \ \ \text{NS} \ . \nn
\eeqa
The coefficients $C_m,  C_{\eta,m}$ are not useful for our purpose. Both series converge when $\epsilon>0$, but the
 limits for $\epsilon\to 0$ are divergent. Extracting the divergent part in the  limit $\epsilon\to 0$, we chose the coefficient $a_m$ in \eqref{eq:Leg} and \eqref{eq:Jac} such that
\beqa
\delta_{\text{reg.}}^m(0)=1 \ \ \text{and} \ \ \delta_{\text{reg.}}^{\eta,m}(0)=1 \ . 
\eeqa
 With this regularisation prescription we obtain the desired result
\beqa                    
\Delta^2_\epsilon(z,w,u,v) \; \buildrel{\hbox{ \text{reg.}}} \over \longrightarrow\; \Delta(z,w)^2 \delta(u-v) \ . \nn
\eeqa
We now define the operators
\beqa
\label{eq:KM-VIr2}
T_\epsilon^a(z,u)&=&\frac \ii 2 M^a_{ij} \noo H_\epsilon^i(z,u) H_\epsilon^j(z,u) \noo\nn\\
L_\epsilon(z,u)&=& \frac 12 \sum \limits_{i=1}^{d} \noo z \frac{\text{d} H_\epsilon^i(z,u)}{\text{d} z} H_\epsilon^i(z,u)\noo + \lambda d
\eeqa
where $\lambda=0$ (resp. $\lambda=\nicefrac{1}{16}$) for fermions with NS (resp. R) boundary conditions in the $z-$direction.
These operators are a direct generalisation  of the Virasoro algebra bilinear in fermions obtained in fermionic strings \cite{go,GSW1}.
Using the Wick theorem together with our regularisation procedure, we carry out computations
similar  as those performed for the two-torus, where  divergences (coming from the double contraction) are regularised by {\bb regulated fermions and the $\zeta-$regularisation}   just described.
We thus obtain at the very end  equations analogous to \eqref{eq:KMVir},  but with $\delta(u-v)$ instead  of $\delta(\theta-\varphi)$.
To extract the modes  we observe the following: for both the NS and the R sectors, the $z-$modes involve integer numbers. As
the functions associated to the Jacobi polynomials $ Q^{(|m-\frac \eta 2|, |m+\frac \eta 2|)}_{\ell-|m|}$ are only
 defined for half-integer values of $m$ and $\ell$, they  are not suitable to expand $T^a(z,u)$ and $L(z,u)$
in the NS sector.  On the other hand, for any $m$, $\{Q_{\ell m}, \ell>|m|\}$ forms a Hilbert basis,  thus any square integrable function on $[-1,1]$
 can be decomposed  with the associated Legendre functions.
Thus
\beqa
T^a(z,u)= \sum \limits_{m \in \mathbb Z} \sum \limits_{\ell > |m|} T^a_{\ell,m} z^{-m} Q_{\ell m}(u) \ \ 
L(z,u)=\sum \limits_{m \in \mathbb Z} \sum \limits_{\ell > |m|} L_{\ell,m} z^{-m} Q_{\ell m}(u)\nn\
\eeqa
or  for both R and NS fermions
\beqa
\label{eq:TaLn2}
T^a_{\ell m}=&\displaystyle  \oint \frac{ \text{d} z z^{m}}{2 \ii \pi z}\int\limits_{-1}^{1} \frac{\text{d} u}{2 } Q_{\ell m}(u) T^a(z,u) \ , \ \ 
L_{\ell m}=  \oint \frac{ \text{d} z\; z^{m}}{2 \ii \pi z}\int\limits_{-1}^{1} \frac{\text{d} u}{2} Q_{\ell m}(u)  L(z,u) \ 
\eeqa
Finally, from \eqref{eq:KMVir} with $\delta(u-v)$ instead of $\delta(\theta-\varphi)$,
standard techniques of Conformal Field Theory,
lead to 
\eqref{eq:mod} with  $c=d/2$ and $k=C_M/2$ for the two-central charges (the $u$ and $v$ integration are trivial). We have thus obtained a fermionic 
 realisation of the semi-direct product  of the Kac-Moody  and Virasoro algebras of the two-sphere
$\hat{\g}(\mathbb S^1 \times \mathbb S^1) \rtimes
\text{Vir}(\mathbb S^1 \times \mathbb S^1)$.  In can be easily verified that, by construction, the corresponding representation obtained acting on the vacuum with creation operators is unitary.\\

\medskip
In this  note we have defined extensions of the Virasoro and Kac-Moody algebras associated to the two-sphere and the
two-torus.  These two algebras  are structurally different. Indeed, the semi-direct product of the Virasoro and the Kac-Moody algebras
of the two-torus contains the usual  semi-direct product of the Virasoro and the Kac-Moody algebras as a subalgebra, whilst  this is not the
case for the two-sphere. This observation comes from the property of the corresponding two-dimensional manifolds, since only $\mathbb S^1\times \mathbb S^1$  arises as a Cartesian product of one-spheres. 
 In particular, for the two-torus, in may be interesting
to consider the limit when one of the  radii of the two-torus  tends to zero, reproducing in this limit the usual Kac-Moody and Virasoro
algebras. In this limit, simultaneously the two-torus `compactifies' to the one-sphere.

\medskip
We have  further  extended  naturally  the usual fermionic realisation of Kac-Moody and Virasoro algebras to their analogue associated with the two-torus and the two-sphere. The crucial point to obtain non-trivial central extensions is the normal ordering prescription, where creation operators are moved to the left of annihilation operators. We have also seen that, in order to regularise infinite divergent sums,  {\bb a regularisation} prescription is needed, basically related to the $\zeta-$regularisation.

\medskip
Finally, it is immediate to observe that the 
 proposed construction can  be easily extended to the $n-$torus, leading to a  hierarchy of embedded algebras, as
 the $n-$torus contains the $(n-1)-$torus as a submanifold.
 In a more general frame, for manifolds ${\cal M}$ that are not of this type, the construction of a fermionic realisation of the Kac-Moody and Virasoro algebras associated to ${\cal M}$
 is technically more delicate, as it requires a careful study of the Hilbert basis and introducing an appropriate regularisation procedure analogous to the regularisation procedure considered in the case of the two-sphere. As shown in \cite{rm}, the construction can be adapted to obtain a bosonic realisation of these algebras in terms of an adapted Vertex operator.

 \bigskip \noindent \textbf{Acknowledgements.}   The authors thank  P. Sorba, A. Marrani, Ch. Schubert and G. 
 Bossard for valuable suggestions and comments. {\bb We are indebted to the anonymous reviewers for many helpful comments, in particular those related to the $\zeta-$regularisation, as well as for pointing out reference \cite{KR}.}
RCS  acknowledges financial support by the research
grant PID2019-106802GB-I00/AEI/10.13039/501100011033 (AEI/ FEDER, UE). 
\bibliographystyle{utphys}

\bibliography{ref-ferm}

\end{document}